\documentclass[%
 reprint,
 amsmath,amssymb,
 aps,
]{revtex4-1}

\usepackage{graphicx}
\usepackage{dcolumn}
\usepackage{bm}
\usepackage{listings}
\usepackage{array,multirow}
\usepackage{caption}
\usepackage{subcaption}
\usepackage[dvipsnames]{xcolor}
\usepackage{colortbl}
\usepackage{adjustbox}
\usepackage{enumitem}  
\usepackage{url}
\usepackage{siunitx}

\usepackage{ctable}

\begin{document}

\title{Binding and excitations in Si$_x$H$_y$ molecular systems using quantum Monte Carlo}

\author{Guangming Wang$^{1}$}
\email{Corresponding author email address: gwang18@ncsu.edu} 
\author{Abdulgani Annaberdiyev$^{1}$}
\author{Lubos Mitas$^{1}$} 

\affiliation{1) Department of Physics, North Carolina State University, Raleigh, North Carolina 27695-8202, USA}


\begin{abstract}
We present high-accuracy correlated calculations of small Si$_x$H$_y$ molecular systems both in the ground and excited states. 
We employ quantum Monte Carlo (QMC) together with a variety of many-body wave function approaches based on basis set expansions. The calculations are carried out in a valence-only framework using recently derived correlation consistent effective core potentials. 
Our primary goal is to understand the fixed-node diffusion QMC errors in both the ground and excited states with single-reference trial wave functions. 
Using a combination of methods, we demonstrate the very high accuracy of the QMC atomization energies being within $\approx$ 0.07 eV or better
when compared with essentially exact results. 
By employing proper choices for trial wave functions, 
we have found that the fixed-node QMC biases for total energies are remarkably uniform ranging between $1-3.5$ \% with absolute values at most $\approx$ 0.2 eV 
across the systems and several types of excitations such as  singlets and triplets as well as low-lying and Rydberg-like states. 
Our results further corroborate that  Si systems, and presumably also related main group IV and V elements of the periodic table (Ge, Sn, etc), exhibit some of the lowest fixed-node biases found in valence-only electronic structure QMC calculations. 
\end{abstract} 


\pacs{Valid PACS appear here}%

\maketitle

\section{Introduction}
\label{section:intro}

Quantum Monte Carlo (QMC) methods belong to a rapidly developing family of many-body approaches that address the central challenge of accurate electronic structure calculations posed by electron-electron correlations. Several flavors of QMC with different sampling strategies are being developed to achieve a better accuracy and use in a wider variety of applications, including model Hamiltonians as well as real molecules and solids. 
Beyond perhaps the most established QMC in real space of particle coordinates \cite{foulkes_quantum_2001,wagner_discovering_2016,becca_quantum_2017},
there are several successful alternative approaches (for example, auxiliary-field QMC, full Configuration Interaction QMC, just to name a few \cite{shee_phaseless_2018, booth_towards_2013}). 
In this work, we use the diffusion Monte Carlo (DMC) algorithm that relies on the fixed-node (FN) approximation to deal with the fermion sign inefficiencies \cite{reynolds_fixednode_1982}.
We simplify the problem by using a valence-only Hamiltonian based on recently derived correlation consistent effective core potentials (ccECPs). Our ccECPs  provide a new level of accuracy and fidelity to the original all-electron Hamiltonian as demonstrated on both atomic and molecular systems \cite{pseudopotential_library,
bennett_new_2017,bennett_new_2018,annaberdiyev_new_2018,wang_new_2019}. 

One of our goals is to test and benchmark these recently established ccECPs  \cite{bennett_new_2017,bennett_new_2018,annaberdiyev_new_2018,wang_new_2019} in realistic systems beyond just the training problems.  This part of work involves also cross-checks with existing 
all-electron calculations that harness state-of-the-art alternative many-body wave function methods. 
The second focal point is to better understand the fixed-node errors in simple molecular systems where high accuracy can be verified by alternative approaches and combined into reference quality data. 
The third goal is to understand the impact of fixed-node errors on excited states which is an area where much less is known overall. 
Indeed, more data on the applicability of QMC methods to excited states of various types (different spin channels, adiabatic vs vertical, Rydberg-like, etc) would be fruitful so as to expand upon a few previous studies \cite{bande_rydberg_2006, schautz_optimized_2004}.


Considering these aims, we have opted for a few molecular Si$_x$H$_y$ systems where extensive basis sets and alternative many-body wave function expansion methods can be used to the full extent. 
The reason for choosing the particular element of
Si was motivated by the fact that its valence electronic structure exhibits some of the smallest fixed-node errors in QMC calculations of the atom, molecule, and also solid systems\cite{rasch_communication_2014}.
Surprisingly, our previous results suggested that this is true even though only single-reference trial wave functions have been employed
\cite{rasch_communication_2014}.
Therefore, another interesting issue we wanted to elucidate was the degree of accuracy that can be achieved for excitations using the same single-reference trial function setting.  
We have included several types of excitations such as
singlets vs triplets, as well as low-lying states in the range of an eV, 
and also very high excitations ($\approx$ 10 eV) with Rydberg-like character. 
The idea behind studying excitations was motivated by an effort to
build a data set that can be used for assessment of fixed-node errors in larger systems where 
expansions in excited states within either Configuration Interaction (CI) or Coupled Cluster (CC) 
methods are very limited or not feasible.

\section{Methods}
\label{section:methods}
We employ our recently generated ccECPs for Si\cite{bennett_new_2018} and H\cite{annaberdiyev_new_2018}.
Previously, extensive ccECP transferability tests on atomic spectra as well as on SiO, Si$_2$, H$_2$ molecular binding energies, equilibrium geometries, and vibrational frequencies have been verified to provide very high accuracy, in some cases significantly beyond widely used ECPs 
of previous generations
\cite{burkatzki_energy-consistent_2007}. 
Our ccECPs and corresponding basis sets are available in Ref. \cite{pseudopotential_library}.

Several quantum chemical methods such as Coupled Cluster with double excitations and perturbative triples (CCSD(T)), as well as triples with perturbative quadruples (CCSDT(Q)), Configuration Interaction with double excitations (CISD), and also triple excitations (CISDT), and finally, Configuration Interaction using a perturbative selection made iteratively (CIPSI) have been used.
We use aug-cc-pV$n$Z basis sets throughout the work since we observed that the augmentations have been important in the description of molecular bonds and some of the excitations.
The basis sets are $\rm{TZ-6Z}$ quality for CCSD(T) and $\rm{DZ-5Z}$ quality for $\rm{CI/CIPSI}$ calculations. 
Using these sets enabled us to recover the total energies and differences within about half of the chemical accuracy threshold (1 kcal/mol $\approx$ 0.0434 eV).
Whenever possible, Hartree-Fock (HF) and correlation energies are extrapolated to complete basis set (CBS) limit as \cite{varandas_accurateab_2007}:
\begin{eqnarray}
    \label{eqn:hf_extrap}
    E^{\rm HF}_n &=& E^{\rm HF}_{\rm CBS} + a \exp\left( - b n \right) \\
    \label{eqn:corr_extrap}
E^{\rm corr}_{n} &=& E^{\rm corr}_{\rm CBS} + \frac{\alpha}{(n+3/8)^{3}} + \frac{\beta}{(n+3/8)^5}
\end{eqnarray}
where $n$ is the basis set cardinal number.
CI/CIPSI energies with multiple roots were also extrapolated with Eqn. \ref{eqn:hf_extrap}.
In some cases, CIPSI energies were extrapolated also in regards to PT2 corrections (using the last 2 data points), shown as exFCI.
CCSDT(Q) calculations were only feasible for small basis sets such as DZ or TZ, therefore, we estimated the CCSDT(Q) CBS limit value by evaluating the energy decrease from CCSD(T) to CCSDT(Q) at a specific basis set and adding that to the CCSD(T) CBS value:
\begin{equation}
    E^{\rm CCSDT(Q)}_{\rm CBS} \approx E^{\rm CCSD(T)}_{\rm CBS} + (E^{\rm CCSDT(Q)}_{n=\rm D/T} - E^{\rm CCSD(T)}_{n=\rm D/T}) 
\end{equation}
This estimation was observed to be reasonable on Si pseudo-atom where CCSDT(Q) was carried out for DZ-6Z basis sets\cite{annaberdiyev_accurate_2020}.
Additionally, Ref.\cite{annaberdiyev_accurate_2020} demonstrates that CCSDT(Q) energies are very close to FCI values (CBS values are same within the fitting errors).


On the QMC side, we have carried out FN-DMC calculations for each system using single-reference and in some cases CI wave functions with an increasing number of determinants (N$_{det}$).
Unless otherwise specified, we used orbitals from density functional theory (DFT)
that are obtained at augmented QZ basis set level. We have tested how the ground and excited states depend on the basis set for DZ-6Z levels and we have found that both states were essentially converged at the QZ basis set level.
We have also tried a doubly-augmented basis set for SiH$_4$ and we did not observe noticeable DMC energy gain for the ground state and for the Rydberg-like singlet excitation.
In all DMC calculations, we use timestep $\tau$ = 0.001 Ha$^{-1}$ with T-moves algorithm \cite{casula_beyond_2006, casula_size-consistent_2010}, which makes the DMC energies rigorously variational. 

For each system, we evaluated the ground state, and the first singlet and triplet excited state energies.
Ground and triplet states can be expressed with a single determinant wavefunction making it suitable to calculate in CC, CI, and QMC methods.
However, for singlet excited states, the correct space-spin symmetry typically leads to a multi-determinant description that is generally challenging for CC methods available in quantum chemistry packages. 
Therefore, for these cases, we have not attempted to calculate the single-reference CC data.
In QMC, it is straightforward to build such wave functions, in the simplest cases as a linear combination of two spin-up$\times$spin-down determinantal products  with a single promoted orbital in one or the other spin channel.
Clearly, it is also straightforward to employ CIS, CISD, or higher-order excitation wave functions in QMC.

In all cases, we have probed for relaxation effects in excited state orbitals when constructing the mentioned two-determinant wave functions. 
To do so, we employed PBE/PBE0 
functionals in GS SCF calculation or even used orbitals from different spin symmetry. 
The most optimal orbitals obtained are indicated where necessary and we also include corresponding HF reference DMC values for comparison.
For the fixed-node DMC and other many-body calculations that use a given set of self-consistent orbitals, we employ notation \mbox{"Method/Orbitals"} throughout the paper.

Note that the DMC method can access the excited states of the given Hamiltonian because the constructed nodal surface constrains it to converge to the ground state of that symmetry. This setting corresponds to all the excitations considered here so that the DMC method provides variational results
\cite{foulkes_quantum_2001}).

All molecular geometries in this work are obtained from experimental values \cite{escribano_absorption_1998, boyd_infrared_1955, gupte_ground_1998}.
Table \ref{mol_geom} provides these geometries.
Software packages \textsc{Molpro}\cite{werner_molpro_2012} and \textsc{Mrcc}\cite{kallay_mrcc_2020} were used for CC calculations. 
\textsc{PySCF}\cite{sun_pyscf_2018}, \textsc{Gamess}\cite{schmidt_general_1993}, and \textsc{Quantum Package}\cite{garniron_quantum_2019} were employed to generate DFT orbtials and CI expansions. 
Subsequent DMC calculations were performed using \textsc{QWalk}\cite{wagner_qwalk_2009} and \textsc{Qmcpack}\cite{kim_qmcpack_2018,kent_qmcpack_2020} packages. 

\begin{table}[!htbp]
\centering
\caption{
Si$_x$H$_y$ molecular geometries used in this work.
Experimental values are used \cite{escribano_absorption_1998, boyd_infrared_1955, gupte_ground_1998}.
All values are in \si\angstrom.
}
\label{mol_geom}
\begin{tabular}{ccrrr}
\hline
\rowcolor{lightgray!35}[2pt] System & Atom & $x$ & $y$ & $z$ \\
SiH$_2$                  & Si &  0.000000 &  0.000000 &  0.000000 \\
(GS \& vertical EX)      & H  &  0.000000 &  0.000000 &  1.514020 \\
                         & H  & -1.513113 &  0.000000 & -0.052390 \\
\hline
SiH$_2$                  & Si &  0.000000 &  0.000000 &  0.000000 \\
(Adiabatic EX $^1$B$_1$) & H  &  0.000000 &  0.000000 &  1.485320 \\
                         & H  & -1.253519 &  0.000000 & -0.796785 \\
\hline
SiH$_2$                  & Si &  0.000000 &  0.000000 &  0.000000 \\
(Adiabatic EX $^3$B$_1$) & H  &  0.000000 &  0.000000 &  1.476800 \\
                         & H  & -1.300289 &  0.000000 & -0.700133 \\
\hline
SiH$_4$                  & Si &  0.0000   &  0.0000   &  0.0000   \\
                         & H  &  0.8544   &  0.8544   &  0.8544   \\
                         & H  & -0.8544   & -0.8544   &  0.8544   \\
                         & H  & -0.8544   &  0.8544   & -0.8544   \\
                         & H  &  0.8544   & -0.8544   & -0.8544   \\
\hline
Si$_2$H$_6$              & Si &  0.0000   &  0.0000   &  1.1600   \\
                         & Si &  0.0000   &  0.0000   & -1.1600   \\
                         & H  &  0.0000   &  1.3865   &  1.6483   \\
                         & H  & -1.2008   & -0.6933   &  1.6483   \\
                         & H  &  1.2008   & -0.6933   &  1.6483   \\
                         & H  &  0.0000   & -1.3865   & -1.6483   \\
                         & H  & -1.2008   &  0.6933   & -1.6483   \\
                         & H  &  1.2008   &  0.6933   & -1.6483   \\
\hline
\end{tabular}
\end{table}

\section{Results and Data}
\label{section:results_data}
\subsection{Total energies for ground states and excitations}

In this subsection, we provide the ground state (GS) and excited state (EX) total energies and excitation gaps using various methods as specified above. 
The results for the Si atom are taken from \cite{annaberdiyev_accurate_2020}.
Table \ref{sih2_gaps} shows the total energies as well as the singlet and triplet gaps of SiH$_2$ molecule where we opted to calculate both vertical and adiabatic excitations since there is a significant geometry relaxation in the adiabatic case according to the experiments\cite{escribano_absorption_1998}.
The total energies and vertical gaps for SiH$_4$ and Si$_2$H$_6$ are listed in Tables \ref{sih4_gaps} and \ref{si2h6_gaps} respectively.

\begin{table*}[!htbp]
\centering
\caption{
SiH$_2$ total energies [Ha] and excitation gaps [eV] using various methods.
All geometries were adopted from experiments \cite{escribano_absorption_1998}.
}
\label{sih2_gaps}
\begin{tabular}{l|l|llll|llll}
\hline
\rowcolor{lightgray!35}[2pt] & \multicolumn{1}{c|}{GS} & \multicolumn{4}{c|}{Vertical excitation} & \multicolumn{4}{c}{Adiabatic excitation} \\
\rowcolor{lightgray!35}[2pt] & \multicolumn{1}{c|}{($^1$A$_1$)} & \multicolumn{2}{c}{$(^1$B$_1$)} & \multicolumn{2}{c|}{$(^3$B$_1$)} & \multicolumn{2}{c}{$(^1$B$_1$)} & \multicolumn{2}{c}{$(^3$B$_1$)} \\
\rowcolor{lightgray!35}[2pt] Method & total[Ha] &  total[Ha] & gap[eV] &  total[Ha] & gap[eV] &  total[Ha] & gap[eV] &  total[Ha] & gap[eV]  \\

CISD/RHF\footnote{CBS values using TZ-5Z extrapolation.}
&    -4.99438  &   -4.83224  &   4.4121  &    -4.95106  &   1.1788  &      -4.84958  &     3.9402  &      -4.96533  &     0.7905  \\
CISDT/RHF\footnote{CBS values using DZ-QZ extrapolation.}
&    -4.99862  &   -4.90946  &   2.4262  &    -4.95583  &   1.1644  &      -4.92695  &     1.9502  &      -4.96979  &     0.7845  \\
exFCI/NatOrb\footnote{N$_{det} \approx 200$ K. Natural orbitals (NatOrb) are obtained from smaller runs with N$_{det} \approx 10$ K. CBS values using DZ-QZ extrapolation.}
&   -5.00676  &  -4.91838  &  2.4051  &   -4.96030  &  1.2642  &     -4.93527  &    1.9453  &     -4.97397  &    0.8924  \\

RCCSD(T)/RHF\footnote{CBS values using TZ-6Z extrapolation.}
&  -5.00627(9) &             &           &  -4.95958(8) &  1.270(3) &                &             &    -4.97339(8) &    0.895(3) \\
CCSDT(Q)/RHF\footnote{Estimated from energy decrease compared to RCCSD(T) at TZ basis set level.}
&  -5.00726(9) &             &           &  -4.96088(8) &  1.262(3) &                &             &    -4.97456(8) &    0.890(3) \\
DMC/RHF
&   -5.0017(1) &  -4.9091(1) &  2.520(4) &   -4.9553(1) &  1.263(4) &     -4.9279(1) &    2.008(4) &     -4.9697(1) &    0.871(4) \\
DMC/PBE\footnote{The lowest GS energy corresponds to DMC/PBE0.}
&   -5.0026(1) &  -4.9136(1) &  2.422(4) &   -4.9579(1) &  1.216(4) &     -4.9317(1) &    1.929(4) &     -4.9718(1) &    0.838(4) \\

& & & & & \\
Yao et al. \footnote{CBS values using DZ-5Z extrapolation with SHCI method from Ref. \cite{yao_almost_2020}.\label{foot:yao}} 
& & & & & & & & &
0.897 \\
Experiment & 
& & 2.36(10)\footnote{Estimated from Fig. 9 in Ref. \cite{duxburyRennerTellerSpin1993}}
& & 1.27(10)$^{\rm g}$ & & 1.92768(1)\footnote{Ref. \cite{escribano_absorption_1998}} 
& & 0.91(4)\footnote{Ref. \cite{balasubramanian_singlettriplet_1986}}, \\
\hline
\end{tabular}
\end{table*}

\begin{table*}[!htbp]
\centering
\caption{
SiH$_4$ total energies [Ha] and vertical excitation gaps [eV] using various methods.
The geometries were obtained from experimental values\cite{boyd_infrared_1955}.
Number of determinants are given in thousands (k).
Note that, counterintuitively, DMC/CIS/PBE ($N_{det}=2.1$ k) gives lower energy than DMC/CIPSI/NatOrb ($N_{det}=360$ k).}
\label{sih4_gaps}
\begin{tabular}{l|r|l|ll|ll}
\hline
\rowcolor{lightgray!35}[2pt] & N$_{det}$ & \multicolumn{1}{c|}{GS ($^1$A$_1$)} & \multicolumn{2}{c|}{EX ($^1$T$_2$)} & \multicolumn{2}{c}{EX ($^3$T$_2$)} \\
\rowcolor{lightgray!35}[2pt] Method & & total[Ha] & total[Ha] & gap[eV] & total[Ha] & gap[eV] \\

CISD/RHF\footnote{CBS values using DZ-QZ extrapolation.} & 
&    -6.2642  &    -5.8231  &   12.003  &    -5.9358  &    8.936  \\
CISD/CAS(8$e^-$, 14o)\footnote{CBS values using TZ-5Z extrapolation.}&  
&    -6.2770  &    -5.9230  &    9.633  &             &           \\
RCCSD(T)/RHF\footnote{CBS values using TZ-6Z extrapolation.} & 
&  -6.2792(1) &             &           &  -5.9563(1) &  8.787(4) \\
CCSDT(Q)/RHF\footnote{Estimated from energy decrease compared to RCCSD(T) at DZ basis set level.} & 
&  -6.2799(1) &             &           &  -5.9599(1) &  8.708(4) \\
DMC/RHF      &
&  -6.2762(1) &  -5.9246(2) &  9.568(6) &  -5.9465(1) &  8.972(4) \\
DMC/PBE\footnote{The EX($^1$T$_2$) state uses EX($^3$T$_2$) UKS/PBE orbitals which resulted in lower energy than using GS orbitals. Spin contamination in UKS was observed to be negligible ($<2S+1>=3.0020831$).}  & 
&  -6.2777(1) &  -5.9301(2) &  9.459(6) &  -5.9532(1) &  8.830(4) \\
VMC/CIS/PBE(aQZ) & 2.1 k 
&  -6.2691(3) &  -5.9210(3) &   9.47(1) &             &           \\
DMC/CIS/PBE(aQZ) & 2.1 k 
&  -6.2767(2) &  -5.9319(2) &  9.382(8) &             &           \\
CIPSI+PT2/PBE(aQZ) &  350 k     
&    -6.2746  &    -5.9224  &    9.584  &             &           \\
VMC/CIPSI/PBE(aQZ) &  350 k
&  -6.2644(2) &  -5.9123(2) &  9.581(8) &             &           \\
CIPSI+PT2/NatOrb(aQZ)\footnote{Natural orbitals (NatOrb) are obtained from a run with N$_{det} \approx 350$ K.} &  360 k  
&    -6.2774  &    -5.9332  &    9.366  &             &           \\
VMC/CIPSI/NatOrb(aQZ)$^{\rm f}$ &  360 k  
&  -6.2695(1) &  -5.9215(1) &  9.470(4) &             &           \\
DMC/CIPSI/NatOrb(aQZ)$^{\rm f}$ &  360 k 
& -6.2774(4) & -5.9302(3) & 9.45(1) & \\

& & & & \\
Lehtonen et al.\footnote{AE CC2 calculation from Ref. \cite{lehtonenCoupledclusterStudiesElectronic2006}}.
& & & & 9.53 & &  \\
Porter et al.\footnote{DMC/CASSCF calculation from Ref. \cite{porter_electronic_2001}}
& & & & 9.44(5) & & 8.89(4) \\
Grossman et al.\footnote{DMC/CASSCF calculation from Ref. \cite{grossman_high_2001}}
& & & & 9.1(1) & & 8.7(1) \\
Experiment \cite{dillon_electron_1985,curtis_low-energy_1989}
& & & & 8.9, 9.7 & & 8.7 \\
Experiment \cite{itohVacuumUltravioletAbsorption1986}
& & & & 9.39(15) \\
Experiment \cite{sutoQuantitativePhotoexcitationStudy1986a}
& & & & 9.43(4)\\
\hline
\end{tabular}
\end{table*}

\begin{table*}[!htbp]
\centering
\caption{
Si$_2$H$_6$ total energies [Ha] and vertical excitation gaps [eV] using various methods.
Experimental geometry was used for this molecule \cite{gupte_ground_1998}.
}
\label{si2h6_gaps}
\begin{tabular}{l|l|ll|ll}
\hline
\rowcolor{lightgray!35}[2pt] & \multicolumn{1}{c|}{GS ($^1$A$_{\rm 1g}$)} & \multicolumn{2}{c|}{EX ($^1$E$_{\rm 1u}$)} & \multicolumn{2}{c}{EX ($^3$A$_{\rm 1g}$)} \\
\rowcolor{lightgray!35}[2pt] Method & total[Ha] & total[Ha] & gap[eV] & total[Ha] & gap[eV] \\

CISD/RHF\footnote{CBS values using DZ-QZ extrapolation.}  
&    -11.3287  &    -10.8183  &   13.889  &    -11.0825  &    6.699  \\
CISD/CAS(14$e^-$, 13o)\footnote{CBS values using DZ-QZ extrapolation.}   
&    -11.3400  &    -11.0623  &    7.557  &              &           \\
RCCSD(T)/RHF\footnote{CBS values using TZ-6Z extrapolation.} 
&  -11.3766(3) &              &           &  -11.1308(5) &   6.69(2) \\
CCSDT(Q)/RHF\footnote{Estimated from energy decrease compared to RCCSD(T) at DZ basis set level.}
&  -11.3782(3) &              &           &  -11.1336(5) &   6.66(2) \\
DMC/RHF
&  -11.3708(2) &  -11.0855(2) &  7.763(8) &  -11.1215(2) &  6.784(8) \\
DMC/PBE\footnote{The lowest EX($^3$A$_{\rm 1g}$) energy corresponds to DMC/PBE0.}   
&  -11.3725(2) &  -11.0934(2) &  7.595(8) &  -11.1248(2) &  6.740(8) \\

& & & \\
Lehtonen et al \footnote{AE CC2 values from \cite{lehtonenCoupledclusterStudiesElectronic2006}.} & & & 7.61 \\
Experiment \cite{dillon_electron_1988}
& & & 7.6 & & $\approx$ 6.7 \footnote{ Experimental value in Ref.[36] of 6.3 eV is probably a mix of vertical and adiabatic excitations.} \\
Experiment \cite{itohVacuumUltravioletAbsorption1986}
& & & 7.56 \\
\hline
\end{tabular}
\end{table*}

Our calculations show solid consistency throughout various methods. 
For each of the molecules, single-reference DMC calculations provide remarkably good accuracy for both the total energies and excitation gaps of singlet and triplet states.
In SiH$_2$ and Si$_2$H$_6$, single-reference DMC excitation gaps also agree very well with experimental values that are known reliably. 
We list CCSD(T) and CCSDT(Q) calculations to illustrate the exact or nearly exact energies obtained by the CBS extrapolations.  
The general agreement between the values from DMC/DFT and CC extrapolations is in the range of or less than $\approx$ 0.1 eV.
For the singlet excitations, DMC/DFT gaps are compared with the CI method with the caveat of less accurate total energies but reasonably good estimations of the gaps. 
In SiH$_2$, we also provide CIPSI+PT2/NatOrb CBS energies with large number of determinants showing that single-reference DMC gaps are accurate to almost the chemical accuracy level. 
In SiH$_4$, we observe higher DMC/PBE excitation gap discrepancies with some experiments for the vertical states. 
Inspection of the experiments \cite{dillon_electron_1985} shows that the singlet peak is very broad and considering the high excitation energies in optical absorption, we conjecture that the measured spectrum might have involved contributions from the adiabatic singlet (which is roughly 0.5 eV lower). 
Here we believe that the genuine vertical excitation is close to 9.4 eV.

Similar considerations
apply to the disilane triplet excitation
where we conjecture that the measured value probably involves 
some degree of adiabacity. Hence, we believe that our values for 
vertical triplet clustered around 6.7 eV represent a valid
prediction for this state.
This requires further study including cleaner and more specific experiments.
Considering the overall accuracy of our calculations and consistency of diversified methods, we assume our result is closer to the true vertical excitation energy. 

\subsection{Binding energies}

Table \ref{mol_binding} presents the Si$_x$H$_y$ computational and experimental atomization energies.
Included are values from extensive CCSD(T) calculations with large basis sets up to 6Z and limited CCSDT(Q) calculations to probe for convergence due to higher-order excitations.
This enabled us to check the extrapolation results more carefully
and provided what we believe are very 
accurate, nearly exact estimations of
the corresponding energy differences. 

Included also are results from published high accuracy studies \cite{feller_theoretical_1999, haunschild_new_2012, martin_accurate_1999} that we use  for comparison and also as a cross-check with all-electron results. 
We find very close agreements with the results of these independent sets of calculations.
Indeed, the differences  are at the level 
of about $\approx$ 0.07 eV or smaller showing thus a very clear consistency 
between different methods.
In fact, the differences between these calculations appear to be overall smaller than are the differences with experiments, see
Tab. \ref{mol_binding}.
Our high accuracy calculations together with results quoted from references, therefore, suggest that experiments mildly underestimate the atomization energies. 
This appears to be true also after all the corrections for core-valence, zero-point motion, and relativity are taken into account as outlined in the references. 

We believe that our results provide a strong argument in favor of the accuracy of the employed ccECPs. 
Below we further analyze the excitations; however, we expect them to be less sensitive to ccECP's quality due to their more delocalized nature and decreased electron density in the core regions. 
Therefore, the main source of errors in excitations is expected to be due to the fixed-node bias.

\begin{table}[!htbp]
\centering
\caption{
Si$_x$H$_y$ atomization energies [eV] using various methods calculated using RCCSD(T), CCSDT(Q), and DMC/DFT methods compared with previous correlated calculations (\cite{feller_theoretical_1999, haunschild_new_2012, greeffQuantumMonteCarlo1997, martin_accurate_1999, porter_electronic_2001}) and with experiments.
The experimental values are corrected to correspond to the bottom of the well (static ions) atomization energies. 
Experimental values of zero-point energy have been used (SiH$_2$: 0.3092 eV, SiH$_4$: 0.8339 eV, Si$_2$H$_6$: 1.3042 eV) \cite{cccbdb}.
}
\label{mol_binding}
\begin{tabular}{clcccc}
\hline
\rowcolor{lightgray!35}[2pt] System & Method & D$_e$ [eV] \\

SiH$_2$      & RCCSD(T) &   6.671(3) \\
             & CCSDT(Q) &   6.673(3) \\
             & DMC/PBE0 &   6.599(4) \\
             & &   \\
             & Feller et al.\footnote{AE UCCSD(T) calculations from Ref. \cite{feller_theoretical_1999} \label{foot:feller}}        & 6.66506 \\
             & Haunschild et al.\footnote{AE frozen core CCSD(T) method calculations from Ref. \cite{haunschild_new_2012} \label{foot:vog}} & 6.665   \\
             & Greef et al.\footnote{DMC calculations from Ref. \cite{greeffQuantumMonteCarlo1997} \label{foot:greef}}               & 6.57(2) \\
             & Yao et al. \footnote{CBS values using DZ-5Z extrapolation with SHCI method\cite{yao_almost_2020}.\label{foot:yao}}  & 6.685 \\
             & Experiments\footnote{Experimental values summarized in Ref. \cite{feller_theoretical_1999} \label{foot:exp}}         & 6.40(9), 6.7(1)\\
\hline
SiH$_4$      & RCCSD(T) &   14.098(3) \\
             & CCSDT(Q) &   14.091(3) \\
             & DMC/PBE  &   14.085(4) \\
             & &   \\
             & Feller et al.\textsuperscript{\ref{foot:feller}}                                                     & 14.0847 \\
             & Haunschild et al.\textsuperscript{\ref{foot:vog}}                                                     & 14.074   \\
             & Greef et al.\textsuperscript{\ref{foot:greef}}                                                       & 14.05(2) \\
             & Martin et al.\footnote{AE CCSD(T) calculation from Ref. \cite{martin_accurate_1999}} & 14.0235 \\
             & Porter et al.\footnote{DMC/CASSCF calculation from Ref. \cite{porter_electronic_2001} \label{foot:port}}    & 14.19(2) \\
             & Yao et al. \textsuperscript{\ref{foot:yao}} & 14.107 \\
             & Experiments\textsuperscript{\ref{foot:exp}}                                                          & 13.96(2), 14.00(2)\\
\hline
Si$_2$H$_6$  & RCCSD(T) & 23.248(9) \\
             & CCSDT(Q) & 23.241(9) \\
             & DMC/PBE  & 23.192(8) \\
             & &   \\
             & Feller et al.\textsuperscript{\ref{foot:feller}}   & 23.2345   \\
             & Haunschild et al.\textsuperscript{\ref{foot:vog}}   & 23.222   \\
             & Greef et al.\textsuperscript{\ref{foot:greef}}     & 23.20(3)  \\
             & Yao et al. \textsuperscript{\ref{foot:yao}} & 23.255 \\
             & Experiments\textsuperscript{\ref{foot:exp}}        & 22.991(4), 23.08(1)   &  \\

\hline
\end{tabular}
\end{table}


\subsection{Fixed-node errors.} 

{\bf Fixed-node biases for  single-reference trial wave functions.}
In order to simplify the analysis 
of nodal biases,
we probed for the impact of basis set sizes using silane as an example (Table \ref{dmc_vs_basis}).  
We found essentially monotonous improvements in total energy with basis set size, although beyond the quadruple zeta basis the improvements have become marginal at 0.1 mHa level. 
Quite surprisingly, a better basis does not automatically imply better nodes (see counter-examples of Bressanini et al.\cite{bressanini_investigation_2005}).
Note that the kinetic energies also follow a similar pattern with monotonous improvements but with slower convergence. 
Also, total energy variances decrease with increasing basis, signalling more accurate trial wave functions.

Let us now analyze the obtained fixed-node errors in more detail (Table \ref{mol_FN}).
In particular, we have identified the following: 

a) As suggested in our previous study, the lowest percentage of fixed-node errors are observed for closed-shell states in  systems with a tetrahedral arrangement of single bonds. 
This holds also in extended systems since almost the same error  is observed in Si crystal
(to be published elsewhere).

b) Somewhat larger errors are observed in open-shell systems and in excitations (Table \ref{mol_FN}) where we see, for example, an increase in percentage by a factor of three between the silane ground and its lowest triplet excited states. 

c) The largest percentage of errors do vary across the systems. For example, the two largest errors are found in the vertical excited singlet state of SiH$_2$ and the vertical excited triplet state of silane.  

Overall, however, it is quite remarkable that in all these systems and states the fixed-node errors are between 1\% and 3.6\%, especially when we consider that single-reference wave functions have been used to approximate the nodal hypersurfaces.

\begin{table*}[!htbp]
\centering
\caption{
Impact of orbitals and basis sets on fixed-node DMC energies [Ha] for SiH$_4$ GS and vertical excitation in aug-cc-pV$n$Z basis set.
Total energies, kinetic energies, and total energy variances are shown for the Slater-Jastrow wave function (single-reference for GS, 2-reference for EX).
}
\label{dmc_vs_basis}
\begin{tabular}{c|ccc|ccc}
\hline
\rowcolor{lightgray!35}[2pt] Basis   & \multicolumn{3}{c|}{GS} & \multicolumn{3}{c}{EX} \\
\rowcolor{lightgray!35}[2pt] $n$Z & Total & Kinetic & Variance & Total & Kinetic & Variance \\
DZ & -6.2740(2) & 3.9521(28) & 0.0589(3) & -5.9229(3) & 3.7568(16) & 0.0559(2) \\
TZ & -6.2765(1) & 3.9658(20) & 0.0507(1) & -5.9252(3) & 3.7651(27) & 0.0468(2) \\
QZ & -6.2774(1) & 3.9690(22) & 0.0428(2) & -5.9277(2) & 3.7650(36) & 0.0392(2) \\
5Z & -6.2776(1) & 3.9771(15) & 0.0365(1) & -5.9279(2) & 3.7683(26) & 0.0346(2) \\
6Z & -6.2777(1) & 3.9779(13) & 0.0355(1) & -5.9272(2) & 3.7625(23) & 0.0330(1) \\
\hline
\end{tabular}
\end{table*}

\begin{table*}[!htbp]
\centering
\caption{
Fixed-node error analysis from single-reference DMC and CC energies [Ha]. For comparison, we have included preliminary results also for Si crystal, to be published elsewhere. MAD denotes mean average deviation.
}
\label{mol_FN}
\begin{tabular}{lllllccc}
\hline
\rowcolor{lightgray!35}[2pt]        &       & DMC/DFT$^{\rm a}$ &  SCF/RHF  &  Estim. Exact  & FN err. &  FN err./Corr. & FN err./N$_{elec}$ \\
\rowcolor{lightgray!35}[2pt] System & State & [Ha] &  [Ha]      &  Corr. [Ha]    & [eV]      &  [\%]            & [eV/electron] \\

           Si &                $^3$P &   -3.7601(1)  &  -3.6724778(1)           &  -0.08957(6) &  0.053(3) &   2.2(1) &  0.0133(8) \\
      SiH$_2$ &            $^1$A$_1$ &   -5.0026(1)  &    -4.85364(8)           &   -0.1536(1) &  0.127(4) &  3.03(9) &  0.0211(6) \\
      SiH$_2$ &      $^1$B$_1$(vert) &   -4.9136(1)  &    -4.77051(7)$^{\rm b}$ &  -0.14787(7) &  0.130(3) &  3.23(7) &  0.0217(5) \\
      SiH$_2$ &      $^1$B$_1$(adia) &   -4.9317(1)  &    -4.79282(7)$^{\rm b}$ &  -0.14245(7) &  0.097(3) &  2.51(7) &  0.0162(5) \\
      SiH$_2$ &      $^3$B$_1$(vert) &   -4.9579(1)  &    -4.83100(7)           &   -0.1299(1) &  0.081(3) &   2.3(1) &  0.0135(6) \\
      SiH$_2$ &      $^3$B$_1$(adia) &   -4.9718(1)  &    -4.84633(7)           &   -0.1282(1) &  0.075(3) &   2.2(1) &  0.0125(6) \\
      SiH$_4$ &            $^1$A$_1$ &   -6.2777(1)  &     -6.0880(1)           &   -0.1919(1) &  0.060(4) &  1.15(7) &  0.0075(5) \\
      SiH$_4$ &            $^3$T$_2$ &   -5.9532(1)  &     -5.7735(1)           &   -0.1864(1) &  0.182(4) &  3.59(7) &  0.0228(5) \\
  Si$_2$H$_6$ &         $^1$A$_{1g}$ &  -11.3725(2)  &    -11.0232(3)           &   -0.3550(4) &   0.16(1) &   1.6(1) &  0.0111(7) \\
  Si$_2$H$_6$ &         $^3$B$_{1g}$ &  -11.1248(2)  &    -10.7878(4)           &   -0.3458(7) &   0.24(1) &   2.5(2) &   0.017(1) \\
\multicolumn{2}{l}{Si cryst/atom$\rm ^c$}  
& & & -0.14342(2) & 0.05(1) & 1.3(3) & 0.013(3) \\
\hline
MAD ground states  &                      &              &                &              &  0.099(3) &  1.99(5) &  0.0132(3) \\
MAD excited states &                      &              &                &              &  0.134(3) &  2.72(4) &  0.0173(3) \\
\hline
\multicolumn{8}{l}{\footnotesize $\rm ^a$ DMC/DFT corresponds to the lowest single-reference DMC energies obtained in this work (2-reference for EX singlet).}  \\
\multicolumn{8}{l}{\footnotesize $\rm ^b$ SiH$_2$ vertical and adiabatic singlet SCF/RHF energies correspond to CASSCF(2$e^-$, 2o) energy with TZ-6Z extrapolations.}  \\
\multicolumn{8}{l}{\footnotesize $\rm ^c$ Assumed  experimental cohesion of 4.68(1) eV/at. and extrapolated DMC/PBE0 value for the Si crystal ground state.} \\
\end{tabular}
\end{table*}

{\bf  Impact of nodal domain topologies vs nodal shapes.} 
It has been conjectured and can be proved in a few cases that the correct number of nodal domains for fermionic ground states is two \cite{mitasStructureFermionNodes2006, ceperleyFermionNodes1991, bajdich_qmc_nodes_2009}.
This implies that all even permutations form a simply connected domain in the space of electron coordinates. In this domain the wave function sign is constant. Similarly, the odd permutations' complementary spatial domain mirrors this with the opposite wave function sign. 
Consequently, in dimensions two and higher, the fermionic ground states generically exhibit a bisection of the particle configuration space with the
boundary given by the node. 
(There are exceptions but a further elaboration on this is out of the scope of this work.) 
For excitations, the situation is a bit more complicated; the nodal count can be two or higher depending on symmetries and other characteristics. 
It is therefore interesting to point out that although the obtained fixed-node biases are rather small, the nodes of most of our calculations with single-reference wave functions are topologically imperfect.
In fact, for most of the states studied here, the count of nodal domains is four. 
This is  due to the single-reference trial wave functions with their anti-symmetric parts given as a product of spin-up times spin-down determinants.

\begin{figure}[!htbp]
\centering
\begin{subfigure}{0.65\columnwidth}
\includegraphics[width=\textwidth]{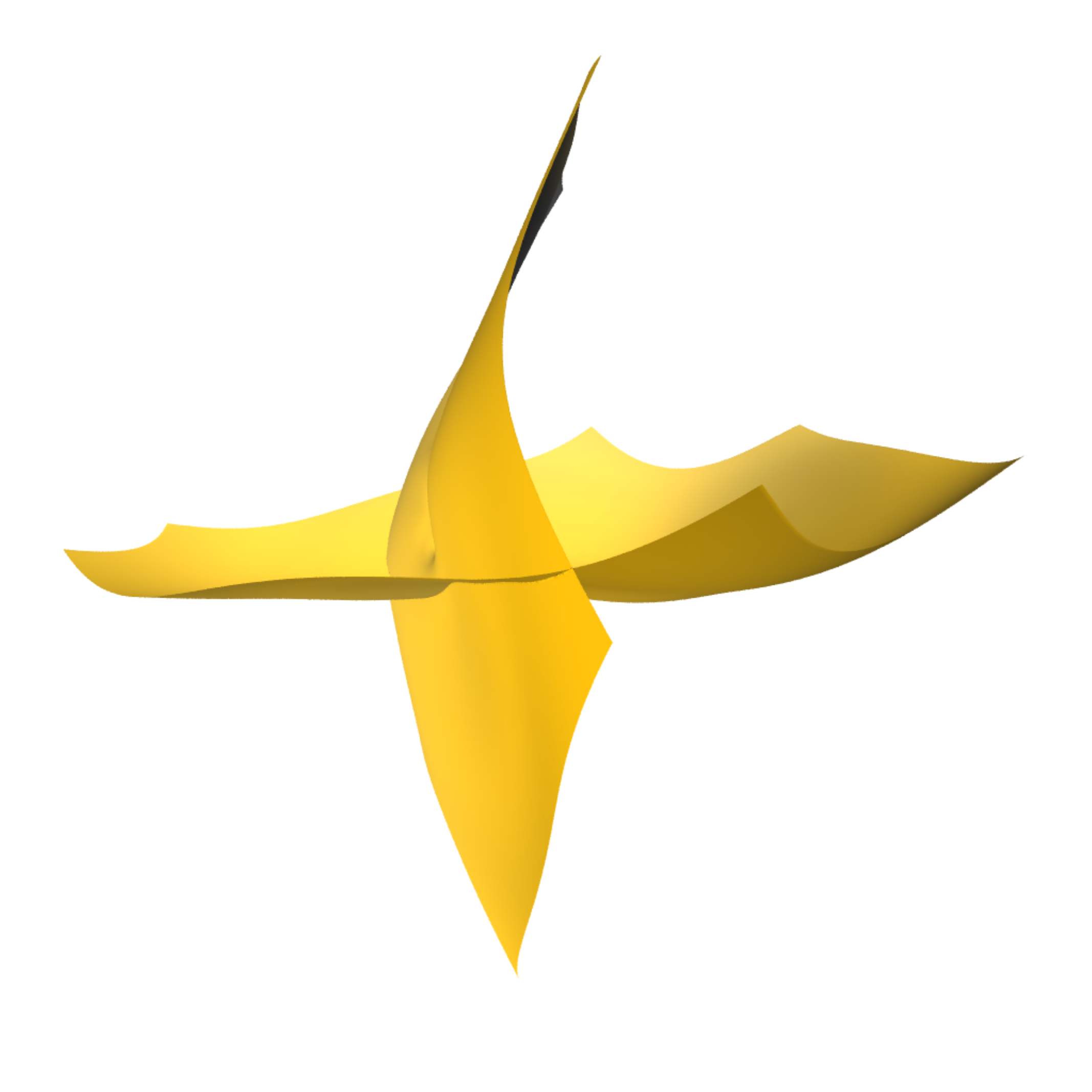}
\caption{}
\end{subfigure}
\begin{subfigure}{0.65\columnwidth}
\includegraphics[width=\textwidth]{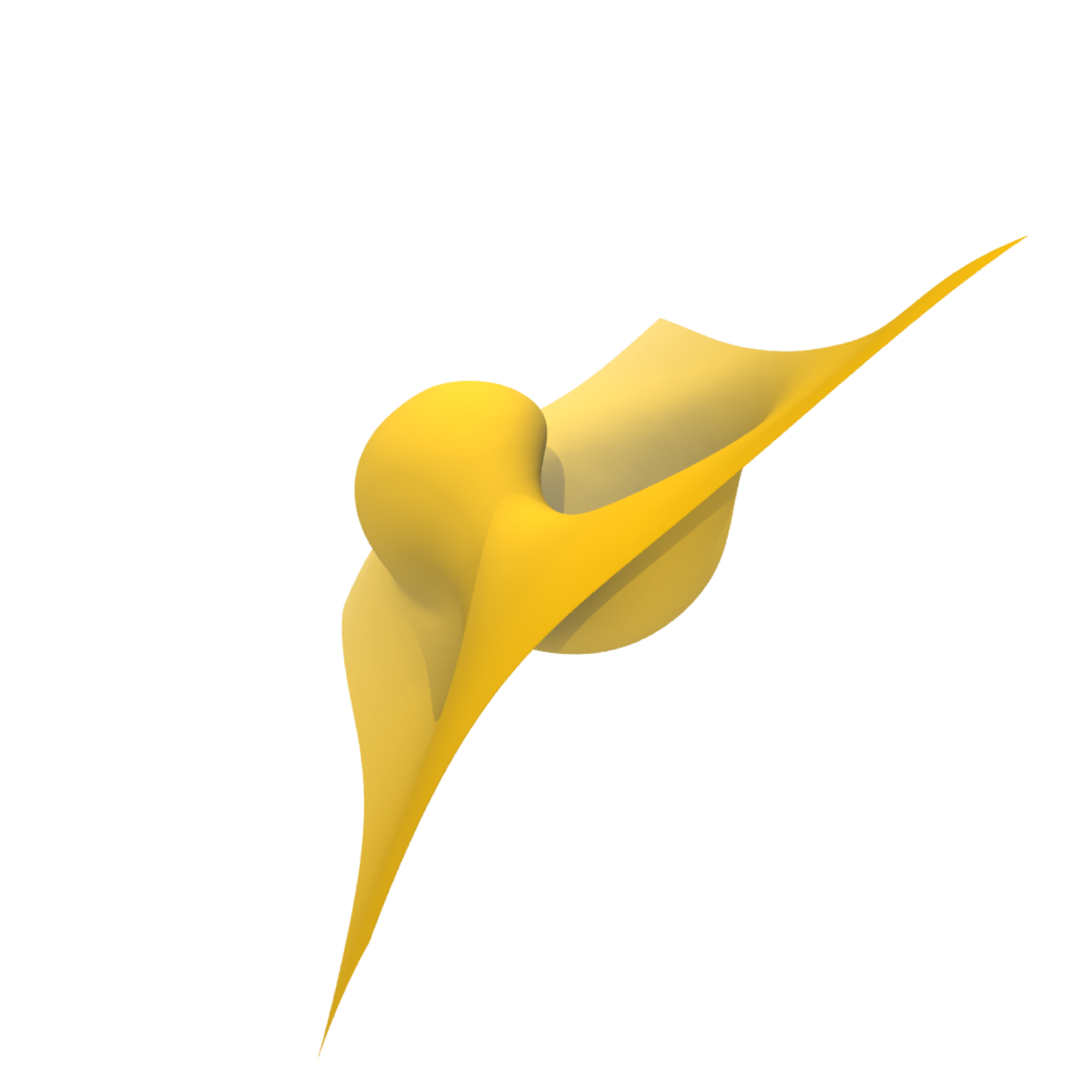}
\caption{}
\end{subfigure}
\caption{
Nodes of SiH$_4$ molecule using 2$e^-$ (up+down electrons at the same point) scan for a) ground state (single-reference), b) excited state singlet (2-reference). Quadrisection of the configurations in case a) and bisection in case b) are clearly visible.
} 
\label{fig:nodes_sih4}
\end{figure}

Interestingly, the following few cases happen to have the correct nodal topologies: 

a) ground state of the Si atom, that has only two nodal domains since there is only one minority spin electron, i.e., the exchange in that channel is absent.

b) SiH$_2$ and SiH$_4$ open-shell  singlet excitations that are given as a linear combination of two determinantal products. 
This form breaks the spin-up and spin-down artificial separation of spatial variables and leads to the fusion of the nodal domains into the minimal two, as it is illustrated in Fig. \ref{fig:nodes_sih4} and can be checked numerically.

It is  rather unexpected that there is no obvious significant difference between the fixed-node biases for correct (2 domains) vs. incorrect (4 domains) topologies. 
Based on these observations, we suppose
that in our systems, the spin-up and -down coordinate subspaces are energetically marginal and the nodal shape in other locations is more important. 
Obviously, with enough variational freedom, the corresponding correct topologies will ultimately emerge as a result of appropriate optimizations.
Unfortunately, our knowledge about the convergence of variational expansions in this respect is rather limited although some promising progress has been achieved very recently with CIPSI methods \cite{dashExcitedStatesSelected2019, dashPerturbativelySelectedConfigurationInteraction2018, caffarelCommunicationImprovedControl2016, caffarelUsingCIPSINodes2016}.
Therefore our results 
can serve as a reference for 
future studies that would address this particular issue.

{\bf Fixed-node biases in excitations.} 
One of our goals was to shed light on how the fixed-node errors impact the excited states. 
Therefore, we have chosen several types of excitations: singlets vs triplets, low-lying states with excited energies on the order of 1 eV vs Rydberg-like states with excitations around 10 eV. 
Interestingly, we do not observe any clear, simple or regular patterns. For SiH$_2$ which has a diradical ground state, we find mild  decreases in the fixed-node biases. 
Presumably, the radical character of the state is still present and therefore the corresponding error persists into excitations with a minor decrease due to orbital restructuring.
On the other hand, for strongly bonded systems with saturated single bonds, the increase is significant, between a factor of two and three.
This is clearly seen in silane that exhibits both the lowest percentage bias (in the ground state) and the highest percentage bias in its excited state.
On an absolute scale, the increases are not substantial, since they are of the order of $0.1 - 0.2$ eV. 
This is quite important for a qualitative assessment of the fixed-node biases in cases where it is, at present, very difficult to estimate their actual values. 
In particular, this is of significant interest  for excitations in periodic (crystal) systems where QMC calculations involve hundreds of valence electrons \cite{gani2020Sisolids}.  

{\bf Multi-reference trial wave functions.} 
We emphasize that it is straightforward to eliminate the fixed-node errors in the studied systems by employing multi-reference trial wave functions. With appropriate effort, significant decrease of the biases can be achieved using CIPSI expansions further boosted through trial wave function re-optimizations. However, our primary goal was to understand the behavior of single-reference fixed-node errors due to our interest in large systems where conclusive, high-accuracy multi-reference study might not be feasible. 

\section{Conclusions}
\label{section:conclusions}

We have presented high accuracy study of Si$_x$H$_y$ systems using both QMC and many-body wave function methods based on expansions in basis sets. We have probed mainly for the level of fixed-node errors in ground states and especially in excited states. We have found 
that the single-reference trial wave functions in FN-DMC provide
excellent descriptions 
of these systems with
1-3.5\% correlation variation in fixed-node biases. 
In general, for excitations these errors were on the larger side, 
but not always and variations in fixed-node biases within the mentioned small range did
not show any clear regular or discernible 
pattern. We have found excellent agreement
with previous independent all-electron state-of-the-art studies thus providing 
a clear support for the accuracy of 
ccECPs used in this study. 
In silane, the vertical singlet excitation appears to be larger in our calculations than in some experiments; 
however, we believe
that new experiments with a more clear 
distinction between vertical and adiabatic
excitations would be very useful. 
Overall, we demonstrate remarkably small
fixed-node biases that further confirm  
our previous finding. The small fixed-node errors are related to particular properties of $s$ and $p$ states in the main group elements of the 2nd row. Similar results can be expected for the same group, such as Ge, Sn, Pb as well as P, As, Sb and Bi
especially in molecules and solids with 
similar single-bond patterns and closed shell 
states.

The obtained data will be useful also for estimations of {\em systematic} errors of QMC results in similar systems with comparable chemical bonding and electronic state calculations.

{\bf Acknowledgements.} 
The presented work used the QWalk package and corresponding tools (55\% of the effort) and it has been supported by the U.S. Department of Energy, Office of Science, Basic Energy Sciences (BES) under the award de-sc0012314. The rest of the work has been using the QMCPACK package (45\% of the effort) and it has been been funded by the U.S. Department of Energy, Office of Science, Basic Energy Sciences, Materials Sciences and Engineering Division,  as  part  of  the  Computational  Materials  Sciences Program and Center for Predictive Simulation of Functional Materials. 
An award of computer time was provided by the Innovative and Novel Computational Impact on Theory and Experiment (INCITE) program. This research used resources of the Argonne Leadership Computing Facility, which is a DOE Office of Science User Facility supported under contract DE-AC02-06CH11357. This research also used resources of the Oak Ridge Leadership Computing Facility, which is a DOE Office of Science User Facility supported under Contract DE-AC05-00OR22725.
The project used also NERSC computational time allocations and resources.
The authors would like to thank 
Anouar Benali and Anthony Scemama for their kind help with
Quantum Package, comments and discussions.

{\bf Data Availability.} 
The input/output files and supporting data generated in this work are published in Materials Data Facility \cite{blaiszik_materials_2016, blaiszik_data_2019} and can be found in Ref. \cite{mdf_data}.

\newpage
\bibliographystyle{unsrt}
\bibliography{main}

\begin{thebibliography}{10}

\bibitem{foulkes_quantum_2001}
W.~M.~C. Foulkes, L.~Mitas, R.~J. Needs, and G.~Rajagopal.
\newblock Quantum {Monte} {Carlo} simulations of solids.
\newblock {\em Reviews of Modern Physics}, 73(1):33--83, January 2001.

\bibitem{wagner_discovering_2016}
Lucas~K Wagner and David~M Ceperley.
\newblock Discovering correlated fermions using quantum {Monte} {Carlo}.
\newblock {\em Reports on Progress in Physics}, 79(9):094501, September 2016.

\bibitem{becca_quantum_2017}
Federico Becca and Sandro Sorella.
\newblock {\em Quantum {Monte} {Carlo} {Approaches} for {Correlated}
  {Systems}}.
\newblock Cambridge University Press, November 2017.

\bibitem{shee_phaseless_2018}
James Shee, Evan~J. Arthur, Shiwei Zhang, David~R. Reichman, and Richard~A.
  Friesner.
\newblock Phaseless {Auxiliary}-{Field} {Quantum} {Monte} {Carlo} on
  {Graphical} {Processing} {Units}.
\newblock {\em J. Chem. Theory Comput.}, 14(8):4109--4121, August 2018.
\newblock Publisher: American Chemical Society.

\bibitem{booth_towards_2013}
George~H. Booth, Andreas Grüneis, Georg Kresse, and Ali Alavi.
\newblock Towards an exact description of electronic wavefunctions in real
  solids.
\newblock {\em Nature}, 493(7432):365--370, January 2013.
\newblock Number: 7432 Publisher: Nature Publishing Group.

\bibitem{reynolds_fixednode_1982}
Peter~J. Reynolds, David~M. Ceperley, Berni~J. Alder, and William~A. Lester.
\newblock Fixed‐node quantum {Monte} {Carlo} for molecules.
\newblock {\em J. Chem. Phys.}, 77(11):5593--5603, December 1982.
\newblock Publisher: American Institute of Physics.

\bibitem{pseudopotential_library}
Pseudopotential {Library}: {A} community website for pseudopotentials/effective
  core potentials developed for high accuracy correlated many-body methods such
  as quantum {Monte} {Carlo} and quantum chemistry.
\newblock \url{https://pseudopotentiallibrary.org}.
\newblock Accessed: 2020-05-30.

\bibitem{bennett_new_2017}
M.~Chandler Bennett, Cody~A. Melton, Abdulgani Annaberdiyev, Guangming Wang,
  Luke Shulenburger, and Lubos Mitas.
\newblock A new generation of effective core potentials for correlated
  calculations.
\newblock {\em J. Chem. Phys.}, 147(22):224106, December 2017.

\bibitem{bennett_new_2018}
M.~Chandler Bennett, Guangming Wang, Abdulgani Annaberdiyev, Cody~A. Melton,
  Luke Shulenburger, and Lubos Mitas.
\newblock A new generation of effective core potentials from correlated
  calculations: 2nd row elements.
\newblock {\em J. Chem. Phys.}, 149(10):104108, September 2018.

\bibitem{annaberdiyev_new_2018}
Abdulgani Annaberdiyev, Guangming Wang, Cody~A. Melton, M.~Chandler Bennett,
  Luke Shulenburger, and Lubos Mitas.
\newblock A new generation of effective core potentials from correlated
  calculations: 3d transition metal series.
\newblock {\em J. Chem. Phys.}, 149(13):134108, October 2018.

\bibitem{wang_new_2019}
Guangming Wang, Abdulgani Annaberdiyev, Cody~A. Melton, M.~Chandler Bennett,
  Luke Shulenburger, and Lubos Mitas.
\newblock A new generation of effective core potentials from correlated
  calculations: 4s and 4p main group elements and first row additions.
\newblock {\em J. Chem. Phys.}, 151(14):144110, October 2019.

\bibitem{bande_rydberg_2006}
Annika Bande, Arne Lüchow, Fabio Della~Sala, and Andreas Görling.
\newblock Rydberg states with quantum {Monte} {Carlo}.
\newblock {\em J. Chem. Phys.}, 124(11):114114, March 2006.
\newblock Publisher: American Institute of Physics.

\bibitem{schautz_optimized_2004}
Friedemann Schautz and Claudia Filippi.
\newblock Optimized {Jastrow}–{Slater} wave functions for ground and excited
  states: {Application} to the lowest states of ethene.
\newblock {\em The Journal of Chemical Physics}, 120(23):10931--10941, June
  2004.

\bibitem{rasch_communication_2014}
Kevin~M. Rasch, Shuming Hu, and Lubos Mitas.
\newblock Communication: {Fixed}-node errors in quantum {Monte} {Carlo}:
  {Interplay} of electron density and node nonlinearities.
\newblock {\em J. Chem. Phys.}, 140(4):041102, January 2014.
\newblock Publisher: American Institute of Physics.

\bibitem{burkatzki_energy-consistent_2007}
M.~Burkatzki, C.~Filippi, and M.~Dolg.
\newblock Energy-consistent pseudopotentials for quantum {Monte} {Carlo}
  calculations.
\newblock {\em The Journal of Chemical Physics}, 126(23):234105, June 2007.

\bibitem{varandas_accurateab_2007}
A.~J.~C. Varandas.
\newblock Accurateab initio-based molecular potentials: from extrapolation
  methods to global modelling.
\newblock {\em Phys. Scr.}, 76(3):C28--C35, August 2007.
\newblock Publisher: IOP Publishing.

\bibitem{annaberdiyev_accurate_2020}
Abdulgani Annaberdiyev, Cody~A. Melton, M.~Chandler Bennett, Guangming Wang,
  and Lubos Mitas.
\newblock Accurate {Atomic} {Correlation} and {Total} {Energies} for
  {Correlation} {Consistent} {Effective} {Core} {Potentials}.
\newblock {\em J. Chem. Theory Comput.}, 16(3):1482--1502, March 2020.
\newblock Publisher: American Chemical Society.

\bibitem{casula_beyond_2006}
Michele Casula.
\newblock Beyond the locality approximation in the standard diffusion {Monte}
  {Carlo} method.
\newblock {\em Phys. Rev. B}, 74(16):161102, October 2006.
\newblock Publisher: American Physical Society.

\bibitem{casula_size-consistent_2010}
Michele Casula, Saverio Moroni, Sandro Sorella, and Claudia Filippi.
\newblock Size-consistent variational approaches to nonlocal pseudopotentials:
  {Standard} and lattice regularized diffusion {Monte} {Carlo} methods
  revisited.
\newblock {\em J. Chem. Phys.}, 132(15):154113, April 2010.
\newblock Publisher: American Institute of Physics.

\bibitem{escribano_absorption_1998}
Rafael Escribano and Alain Campargue.
\newblock Absorption spectroscopy of {SiH2} near 640 nm.
\newblock {\em J. Chem. Phys.}, 108(15):6249--6257, April 1998.
\newblock Publisher: American Institute of Physics.

\bibitem{boyd_infrared_1955}
D.~R.~J. Boyd.
\newblock Infrared {Spectrum} of {Trideuterosilane} and the {Structure} of the
  {Silane} {Molecule}.
\newblock {\em J. Chem. Phys.}, 23(5):922--926, May 1955.
\newblock Publisher: American Institute of Physics.

\bibitem{gupte_ground_1998}
Girish~R. Gupte and R.~Prasad.
\newblock Ground {State} {Geometries} and {Vibrational} {Spectra} of {Small}
  {Hydrogenated} {Silicon} {Clusters} using {Nonorthogonal} {Tight}-{Binding}
  {Molecular} {Dynamics}.
\newblock {\em Int. J. Mod. Phys. B}, 12(15):1607--1622, June 1998.
\newblock Publisher: World Scientific Publishing Co.

\bibitem{werner_molpro_2012}
Hans-Joachim Werner, Peter~J. Knowles, Gerald Knizia, Frederick~R. Manby, and
  Martin Schütz.
\newblock Molpro: a general-purpose quantum chemistry program package.
\newblock {\em WIREs Computational Molecular Science}, 2(2):242--253, 2012.

\bibitem{kallay_mrcc_2020}
Mihály Kállay, Péter~R. Nagy, Dávid Mester, Zoltán Rolik, Gyula Samu,
  József Csontos, József Csóka, P.~Bernát Szabó, László Gyevi-Nagy,
  Bence Hégely, István Ladjánszki, Lóránt Szegedy, Bence Ladóczki, Klára
  Petrov, Máté Farkas, Pál~D. Mezei, and Ádám Ganyecz.
\newblock The {MRCC} program system: {Accurate} quantum chemistry from water to
  proteins.
\newblock {\em J. Chem. Phys.}, 152(7):074107, February 2020.
\newblock Publisher: American Institute of Physics.

\bibitem{sun_pyscf_2018}
Qiming Sun, Timothy~C. Berkelbach, Nick~S. Blunt, George~H. Booth, Sheng Guo,
  Zhendong Li, Junzi Liu, James~D. McClain, Elvira~R. Sayfutyarova, Sandeep
  Sharma, Sebastian Wouters, and Garnet Kin-Lic Chan.
\newblock {PySCF}: the {Python}-based simulations of chemistry framework.
\newblock {\em WIREs Computational Molecular Science}, 8(1):e1340, 2018.

\bibitem{schmidt_general_1993}
Michael~W. Schmidt, Kim~K. Baldridge, Jerry~A. Boatz, Steven~T. Elbert, Mark~S.
  Gordon, Jan~H. Jensen, Shiro Koseki, Nikita Matsunaga, Kiet~A. Nguyen, Shujun
  Su, Theresa~L. Windus, Michel Dupuis, and John~A. Montgomery.
\newblock General atomic and molecular electronic structure system.
\newblock {\em Journal of Computational Chemistry}, 14(11):1347--1363, 1993.

\bibitem{garniron_quantum_2019}
Yann Garniron, Thomas Applencourt, Kevin Gasperich, Anouar Benali, Anthony
  Ferté, Julien Paquier, Barthélémy Pradines, Roland Assaraf, Peter
  Reinhardt, Julien Toulouse, Pierrette Barbaresco, Nicolas Renon, Grégoire
  David, Jean-Paul Malrieu, Mickaël Véril, Michel Caffarel, Pierre-François
  Loos, Emmanuel Giner, and Anthony Scemama.
\newblock Quantum {Package} 2.0: {An} {Open}-{Source} {Determinant}-{Driven}
  {Suite} of {Programs}.
\newblock {\em J. Chem. Theory Comput.}, 15(6):3591--3609, June 2019.
\newblock Publisher: American Chemical Society.

\bibitem{wagner_qwalk_2009}
Lucas~K. Wagner, Michal Bajdich, and Lubos Mitas.
\newblock {QWalk}: {A} quantum {Monte} {Carlo} program for electronic
  structure.
\newblock {\em Journal of Computational Physics}, 228(9):3390--3404, May 2009.

\bibitem{kim_qmcpack_2018}
Jeongnim Kim, Andrew~D. Baczewski, Todd~D. Beaudet, Anouar Benali, M.~Chandler
  Bennett, Mark~A. Berrill, Nick~S. Blunt, Edgar Josué~Landinez Borda, Michele
  Casula, David~M. Ceperley, Simone Chiesa, Bryan~K. Clark, Raymond~C. Clay,
  Kris~T. Delaney, Mark Dewing, Kenneth~P. Esler, Hongxia Hao, Olle Heinonen,
  Paul R.~C. Kent, Jaron~T. Krogel, Ilkka Kylänpää, Ying~Wai Li, M.~Graham
  Lopez, Ye~Luo, Fionn~D. Malone, Richard~M. Martin, Amrita Mathuriya, Jeremy
  McMinis, Cody~A. Melton, Lubos Mitas, Miguel~A. Morales, Eric Neuscamman,
  William~D. Parker, Sergio D.~Pineda Flores, Nichols~A. Romero, Brenda~M.
  Rubenstein, Jacqueline A.~R. Shea, Hyeondeok Shin, Luke Shulenburger,
  Andreas~F. Tillack, Joshua~P. Townsend, Norm~M. Tubman, Brett Van~Der Goetz,
  Jordan~E. Vincent, D.~ChangMo Yang, Yubo Yang, Shuai Zhang, and Luning Zhao.
\newblock {QMCPACK}: an open sourceab initioquantum {Monte} {Carlo} package for
  the electronic structure of atoms, molecules and solids.
\newblock {\em J. Phys.: Condens. Matter}, 30(19):195901, April 2018.
\newblock Publisher: IOP Publishing.

\bibitem{kent_qmcpack_2020}
P.~R.~C. Kent, Abdulgani Annaberdiyev, Anouar Benali, M.~Chandler Bennett,
  Edgar~Josué Landinez~Borda, Peter Doak, Hongxia Hao, Kenneth~D. Jordan,
  Jaron~T. Krogel, Ilkka Kylänpää, Joonho Lee, Ye~Luo, Fionn~D. Malone,
  Cody~A. Melton, Lubos Mitas, Miguel~A. Morales, Eric Neuscamman, Fernando~A.
  Reboredo, Brenda Rubenstein, Kayahan Saritas, Shiv Upadhyay, Guangming Wang,
  Shuai Zhang, and Luning Zhao.
\newblock {QMCPACK}: {Advances} in the development, efficiency, and application
  of auxiliary field and real-space variational and diffusion quantum {Monte}
  {Carlo}.
\newblock {\em J. Chem. Phys.}, 152(17):174105, May 2020.
\newblock Publisher: American Institute of Physics.

\bibitem{yao_almost_2020}
Yuan Yao, Emmanuel Giner, Junhao Li, Julien Toulouse, and C.~J. Umrigar.
\newblock Almost exact energies for the {Gaussian}-2 set with the
  semistochastic heat-bath configuration interaction method.
\newblock {\em arXiv:2004.10059 [physics, physics:quant-ph]}, June 2020.
\newblock arXiv: 2004.10059.

\bibitem{duxburyRennerTellerSpin1993}
Geoffrey Duxbury, Alexander Alijah, and Reuben~R. Trieling.
\newblock Renner\textendash{{Teller}} and spin\textendash orbit interactions in
  {{SiH2}}.
\newblock {\em J. Chem. Phys.}, 98(2):811--825, January 1993.

\bibitem{balasubramanian_singlettriplet_1986}
K.~Balasubramanian and A.~D. McLean.
\newblock The singlet–triplet energy separation in silylene.
\newblock {\em The Journal of Chemical Physics}, 85(9):5117--5119, November
  1986.

\bibitem{lehtonenCoupledclusterStudiesElectronic2006}
Olli Lehtonen and Dage Sundholm.
\newblock Coupled-cluster studies of the electronic excitation spectra of
  silanes.
\newblock {\em J. Chem. Phys.}, 125(14):144314, October 2006.

\bibitem{porter_electronic_2001}
A.~R. Porter, O.~K. Al-Mushadani, M.~D. Towler, and R.~J. Needs.
\newblock Electronic excited-state wave functions for quantum {Monte} {Carlo}:
  {Application} to silane and methane.
\newblock {\em J. Chem. Phys.}, 114(18):7795--7804, April 2001.
\newblock Publisher: American Institute of Physics.

\bibitem{grossman_high_2001}
Jeffrey~C. Grossman, Michael Rohlfing, Lubos Mitas, Steven~G. Louie, and
  Marvin~L. Cohen.
\newblock High {Accuracy} {Many}-{Body} {Calculational} {Approaches} for
  {Excitations} in {Molecules}.
\newblock {\em Phys. Rev. Lett.}, 86(3):472--475, January 2001.
\newblock Publisher: American Physical Society.

\bibitem{dillon_electron_1985}
Michael~A. Dillon, R.‐G. Wang, Z.‐W. Wang, and David Spence.
\newblock Electron impact spectroscopy of silane and germane.
\newblock {\em J. Chem. Phys.}, 82(7):2909--2917, April 1985.
\newblock Publisher: American Institute of Physics.

\bibitem{curtis_low-energy_1989}
Martin~G. Curtis and Isobel~C. Walker.
\newblock Low-energy electron-impact excitation of methane, silane,
  tetrafluoromethane and tetrafluorosilane.
\newblock {\em Journal of the Chemical Society, Faraday Transactions 2},
  85(6):659, 1989.

\bibitem{itohVacuumUltravioletAbsorption1986}
Uichi Itoh, Yasutake Toyoshima, Hideo Onuki, Nobuaki Washida, and Toshio Ibuki.
\newblock Vacuum ultraviolet absorption cross sections of {{SiH4}}, {{GeH4}},
  {{Si2H6}}, and {{Si3H8}}.
\newblock {\em J. Chem. Phys.}, 85(9):4867--4872, November 1986.

\bibitem{sutoQuantitativePhotoexcitationStudy1986a}
Masako Suto and L.~C. Lee.
\newblock Quantitative photoexcitation study of {{SiH4}} in vacuum ultraviolet.
\newblock {\em J. Chem. Phys.}, 84(3):1160--1164, February 1986.

\bibitem{dillon_electron_1988}
Michael~A. Dillon, David Spence, Ludwig Boesten, and Hiroshi Tanaka.
\newblock Electron energy loss spectroscopy of disilane.
\newblock {\em The Journal of Chemical Physics}, 88(7):4320--4323, April 1988.

\bibitem{feller_theoretical_1999}
David Feller and David~A. Dixon.
\newblock Theoretical {Study} of the {Heats} of {Formation} of {Small}
  {Silicon}-{Containing} {Compounds}.
\newblock {\em J. Phys. Chem. A}, 103(32):6413--6419, August 1999.
\newblock Publisher: American Chemical Society.

\bibitem{haunschild_new_2012}
Robin Haunschild and Wim Klopper.
\newblock New accurate reference energies for the {G2}/97 test set.
\newblock {\em The Journal of Chemical Physics}, 136(16):164102, April 2012.

\bibitem{martin_accurate_1999}
Jan M.~L. Martin, Kim~K. Baldridge, and Timothy~J. Lee.
\newblock Accurate \textit{ab initio} anharmonic force field and heat of
  formation for silane.
\newblock {\em Molecular Physics}, 97(8):945--953, October 1999.

\bibitem{greeffQuantumMonteCarlo1997}
C.~W. Greeff and Jr. Lester, W.~A.
\newblock Quantum {{Monte Carlo}} binding energies for silicon hydrides.
\newblock {\em J. Chem. Phys.}, 106(15):6412--6417, April 1997.

\bibitem{cccbdb}
{C}omputational {C}hemistry {C}omparison and {B}enchmark {D}atabase, {N}ational
  {I}nstitute of {S}tandards and {T}echnology.
\newblock \url{https://cccbdb.nist.gov/exp1x.asp}.
\newblock Accessed: 2020-07-19.

\bibitem{bressanini_investigation_2005}
Dario Bressanini, Gabriele Morosi, and Silvia Tarasco.
\newblock An investigation of nodal structures and the construction of trial
  wave functions.
\newblock {\em The Journal of Chemical Physics}, 123(20):204109, November 2005.

\bibitem{mitasStructureFermionNodes2006}
Lubos Mitas.
\newblock Structure of {{Fermion Nodes}} and {{Nodal Cells}}.
\newblock {\em Phys. Rev. Lett.}, 96(24):240402, June 2006.

\bibitem{ceperleyFermionNodes1991}
D.~M. Ceperley.
\newblock Fermion nodes.
\newblock {\em J Stat Phys}, 63(5):1237--1267, June 1991.

\bibitem{bajdich_qmc_nodes_2009}
M.~Bajdich and L~Mitas.
\newblock {E}lectronic structure quantum {M}onte {C}arlo.
\newblock {\em Acta Physica Slovaca}, 59(2):81--168, May 2009.

\bibitem{dashExcitedStatesSelected2019}
Monika Dash, Jonas Feldt, Saverio Moroni, Anthony Scemama, and Claudia Filippi.
\newblock Excited {{States}} with {{Selected Configuration
  Interaction}}-{{Quantum Monte Carlo}}: {{Chemically Accurate Excitation
  Energies}} and {{Geometries}}.
\newblock {\em J. Chem. Theory Comput.}, 15(9):4896--4906, September 2019.

\bibitem{dashPerturbativelySelectedConfigurationInteraction2018}
Monika Dash, Saverio Moroni, Anthony Scemama, and Claudia Filippi.
\newblock Perturbatively {{Selected Configuration}}-{{Interaction Wave
  Functions}} for {{Efficient Geometry Optimization}} in {{Quantum Monte
  Carlo}}.
\newblock {\em J. Chem. Theory Comput.}, 14(8):4176--4182, August 2018.

\bibitem{caffarelCommunicationImprovedControl2016}
Michel Caffarel, Thomas Applencourt, Emmanuel Giner, and Anthony Scemama.
\newblock Communication: {{Toward}} an improved control of the fixed-node error
  in quantum {{Monte Carlo}}: {{The}} case of the water molecule.
\newblock {\em J. Chem. Phys.}, 144(15):151103, April 2016.

\bibitem{caffarelUsingCIPSINodes2016}
Michel Caffarel, Thomas Applencourt, Emmanuel Giner, and Anthony Scemama.
\newblock Using {{CIPSI Nodes}} in {{Diffusion Monte Carlo}}.
\newblock In {\em Recent {{Progress}} in {{Quantum Monte Carlo}}}, volume 1234
  of {\em {{ACS Symposium Series}}}, chapter~2, pages 15--46. {American
  Chemical Society}, January 2016.

\bibitem{gani2020Sisolids}
Future work to be published on {Si} crystal using {Q}{M}{C}.

\bibitem{blaiszik_materials_2016}
B.~Blaiszik, K.~Chard, J.~Pruyne, R.~Ananthakrishnan, S.~Tuecke, and I.~Foster.
\newblock The {Materials} {Data} {Facility}: {Data} {Services} to {Advance}
  {Materials} {Science} {Research}.
\newblock {\em JOM}, 68(8):2045--2052, August 2016.

\bibitem{blaiszik_data_2019}
Ben Blaiszik, Logan Ward, Marcus Schwarting, Jonathon Gaff, Ryan Chard, Daniel
  Pike, Kyle Chard, and Ian Foster.
\newblock A {Data} {Ecosystem} to {Support} {Machine} {Learning} in {Materials}
  {Science}.
\newblock {\em MRC}, 9(4):1125--1133, December 2019.
\newblock arXiv: 1904.10423.

\bibitem{mdf_data}
Guangming Wang, Abdulgani Annaberdiyev, and Lubos Mitas.
\newblock Dataset for ``{B}inding and excitations in {S}i$_x${H}$_y$ molecular
  systems using quantum {M}onte {C}arlo''.
\newblock \mbox{\url{https://doi.org/10.18126/dxpz-vo2i}}, 2020.

\end{thebibliography}

\end{document}